\documentclass[twocolumn]{aastex62}

\graphicspath{{./}{figures/}}

\received{X}
\revised{Y}
\accepted{Z}
\submitjournal{AAS Journals}

\shorttitle{Vides et al.}
\shortauthors{Vides et al.}

\begin{document}

\title{Model of the Search For Extraterrestrial Intelligence with Coronagraphic Imaging}

\correspondingauthor{Christina L. Vides}
\email{clvides@cpp.edu}

\author[0000-0002-0786-7307]{Christina L. Vides}
\affil{California State Polytechnic University, Pomona, 3801 W Temple Ave., Pomona, CA, 91750 USA}

\author{Bruce Macintosh}
\affil{Stanford University, 450 Serra Mall, Stanford, CA 94305, USA}

\author[0000-0002-4955-0471]{Breanna A. Binder}
\affil{California State Polytechnic University, Pomona, 3801 W Temple Ave., Pomona, CA, 91750 USA}

\author[0000-0002-4918-0247]{Robert J. De Rosa}
\affil{Stanford University, 450 Serra Mall, Stanford, CA 94305, USA}

\author{Jean-Baptiste Ruffio}
\affil{Stanford University, 450 Serra Mall, Stanford, CA 94305, USA}

\author{Dmitry Savransky}
\affiliation{Cornell University, 616-A Space Sciences Building, Ithaca, NY 14853, USA}

\begin{abstract}
We present modeled detection limits of the Gemini Planet Imager (GPI) and the Wide-Field Infrared Space Telescope (WFIRST) to an optical and infrared laser which could be used by an extraterrestrial civilization to signal their presence. GPI and WFIRST could utilize a coronagraph to search for extraterrestrial intelligence (SETI) in the present and future. We use archival data for GPI stars and simulated WFIRST observations to find the detectable flux ratio of a laser signal to residual scattered starlight around the target star. This flux ratio is then converted to detectable power as a function of distance from the parent star. For GPI, we assume a monochromatic laser wavelength of 1.55 \micron\, and a wavelength of 575 nm  for WFIRST. We assume the lasers are projected through a 10-m aperture, and that the intensity of the laser beam follows a Gaussian profile. Our analysis is performed on 6 stars with spectral types later than F within 20 pc (with an emphasis on solar analogs at different distances). The most notable result is the detection limit for \(\tau\) Ceti, a G5V star with four known exoplanets, two of those within the habitable zone (HZ). The result shows that a 24 kW laser is detectable from \(\tau\) Ceti from outside of the HZ with GPI and a 7.3 W laser is detectable from within \(\tau\) Ceti's HZ by WFIRST.  

\end{abstract}


\section{Introduction} \label{sec:intro}
\subsection{SETI} \label{subsec:intro_SETI}
	The search for extraterrestrial intelligence (SETI) has traditionally been carried out in radio wavelengths, beginning with the suggestion of \citet{Cocconi+59} to monitor wavelengths near the 21 cm emission lines of hydrogen. Radio electromagnetic (EM) radiation could be an ideal mode of communication between us and extraterrestrial intelligence (ETI) since decades-old technology allows for a large field of view and high sensitivity \citep{Deboer06}. While radio SETI has advantages, due to null results and a rapid increase in technology and telescope sensitivities, innovative suggestions on how to conduct SETI have emerged. One such method is optical and infrared SETI (OIRSETI) which allows for the potential of detecting technosignatures from an ET civilization. Such technosignatures could include city lights, megastructure transits, and atmospheric pollutants, but also deliberately transmitted laser signals \citep{Maire+16b,Wright+18}. Searching for laser transmitters was first suggested by \citet{Townes+61} and has been the major focus of OIRSETI. Since dispersion effects from interstellar medium are considered to be negligible at the optical and near infrared wavelengths, one could imagine that if ETI would want to intentionally signal their existence, these wavelengths would be appealing \citep{Townes83, Howard+04,Horowitz+01}.  Any hypothetical laser will have much lower total flux than a planet-hosting star, so some method is needed to distinguish the laser signal from the star. This can be done temporally, searching for photon arrival time coincidences in a pulsed laser \citep{Wright+14,Wright+04,Howard+04}, or spectroscopically, using high spectral resolution to distinguish a narrow-band signal of a pulsed or continuous wave (CV) laser from the stellar continuum \citep[][]{Tellis+15,Tellis,Reines+02}.  
	
	In some ways, distinguishing a faint laser signal from a bright planetary signal is similar to the problem of detecting a faint planet near a bright star. Just as in planet detection, different methods have their own advantages and disadvantages. One planet-detection method is direct imaging -- spatially resolving the stellar and planetary signal from each other \citep{Marois+08,Macintosh+14,Mennesson+18} -- usually with a coronagraph. If a laser transmitter is spatially separated from the star it orbits, direct imaging could in principal detect and distinguish the laser light being transmitted to Earth from the starlight.

\subsection{Direct Imaging with a Coronagraph}

\label{subsec:SETI}
    Coronagraphic exoplanet direct imaging technology has progressed rapidly in the past decade with advanced adaptive optics, novel starlight suppression systems and signal processing methods \citep{Bowler16}. Ground-based coronagraphs routinely image young giant planets at planet to star flux ratios up to  $10^{-6}$. The Gemini Planet Imager \citep[GPI;][]{Macintosh+14} is an example of such an instrument. With suitable image processing and target selection, it would also be capable of detecting OIRSETI signals. 
    
    The ultimate goal of coronagraphic science is detection of mature planets, preferably Earth-like, at flux ratios of $10^{-10}$. Achieving that sensitivity requires a space-based coronagraph combining starlight suppression and active-optics deformable mirrors. The first such mission to fly will likely be the Wide-Field Infrared Survey \citep[WFIRST,][]{Akeson+19}. This is a proposed  multipurpose space mission carrying a technology-demonstrated coronagraph, set for launch in the mid 2020's, which will be more sensitive than GPI by a factor of $\sim10^4$. 
    
    In coronagraphic SETI (CORSETI), a telescope would image a nearby star to search for laser signals. Appropriate targets would be relatively nearby stars (to maximize the angular separation between laser and star) old enough to have developed a technological civilization, which favors stars of spectral type F, G, K, and even M. Moderately-long (minutes to hours) coronagraphic images would be searched for bright point sources. A laser signal could be distinguished from a planet through its spectral signature or temporal variability.  Unlike fast-photometry OIRSETI experiments, but similar to high-resolution spectroscopic measurements, CORSETI would be sensitive to both pulsed and CW lasers. A coronagraph could also be combined with classical OIRSETI techniques using high temporal or spectral resolution \citep{wang_ji2017}. This would further enhance the ability to distinguish between residual starlight and a true source, but we will not model the limits here.
   
   The intent of this paper is to determine the effectiveness of GPI and WFIRST at detecting such lasers. We calculate the detection limits of each instrument for a sample of nearby and Sun-like target stars at a range of separations. Since laser transmitters might be co-located with an inhabited planet, we also evaluate the spatial abilities of each instrument in imaging the habitable zone (HZ) of a key SETI target star, \(\tau\) Ceti, which has a possible detection of a super-Earth planet in its habitable zone, and determine the detection limit in this region.

Although a laser is of course much less luminous than a planet, we might still expect high-performance coronagraphs to be able to detect a laser signal.  This can be demonstrated by calculating the ratio of the luminosity of the Earth (approximately is \(10^{16}\) W) to the solid angle over which the Earth radiates (4\(\pi\) sr) and the ratio of a 10 W laser to the solid angle of the beam transmitted from it. The latter is approximately the diffraction limit of the launch telescope, which is  the ratio of the laser wavelength over the diameter of a transmitter. If we considering a 10 meter transmitter and a laser wavelength of 1 \micron, the ratios of both the luminosity of the Earth/emitted solid angle and laser to transmitted solid angle are both approximately \(10^{15}\) W/sr. All of the laser's energy is concentrated into a small beam of \(10^{-14}\) sr, whereas the energy of the Earth is  radial normal to its surface, therefore it is conceivable that the laser could outshine a planet when observed at the appropriate wavelength. 

\begin{figure*}
    \centering
    \begin{tabular}{cc}
         \includegraphics[height=2.5in]{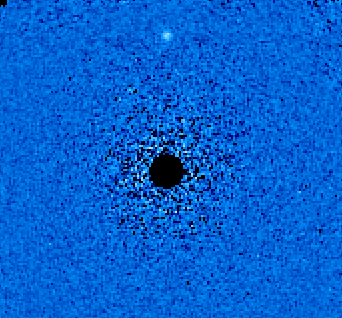} & 
         \includegraphics[height=2.85in, width=4.5in]{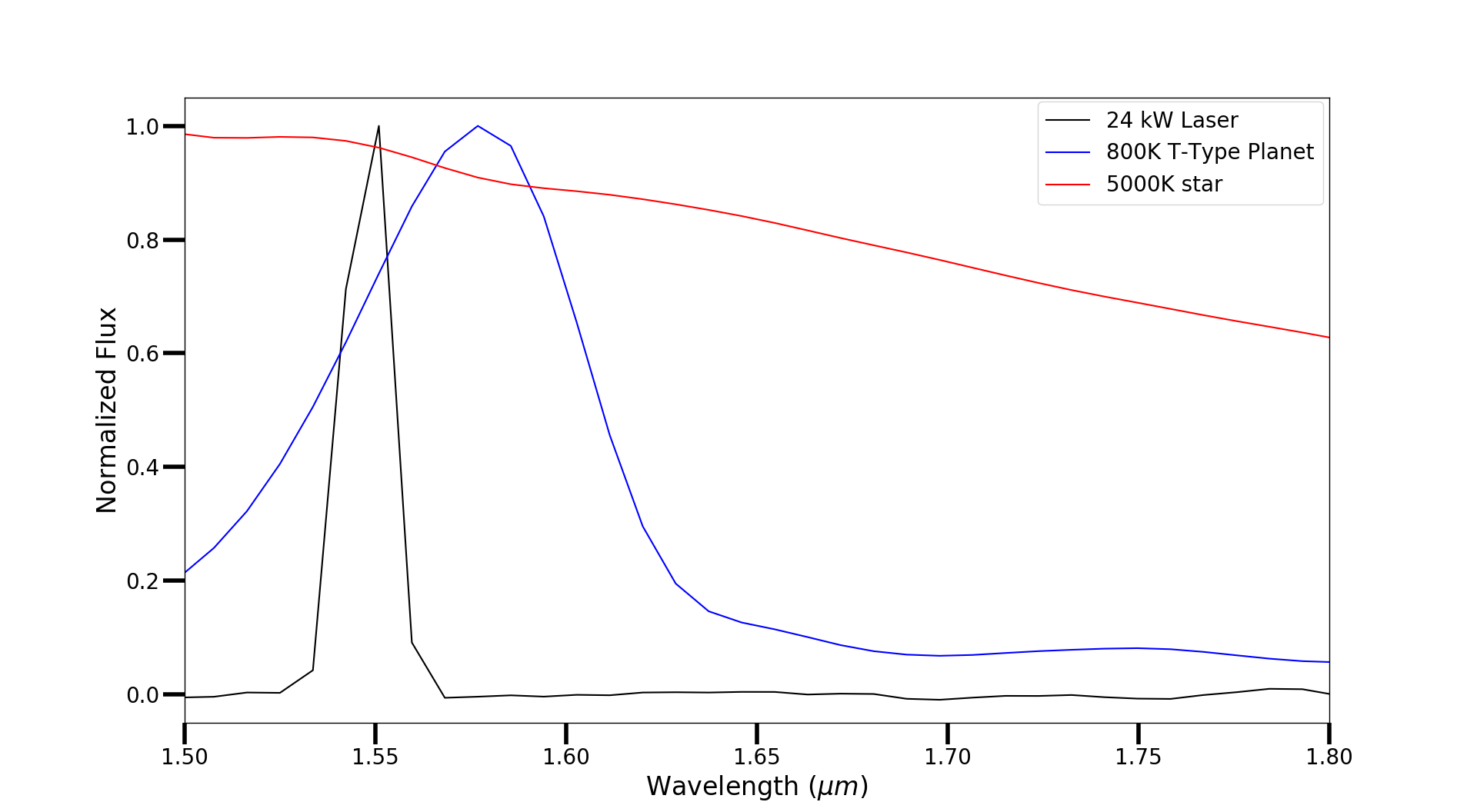}  \\
    \end{tabular}
    \caption{A laser spectra injected into a data cube and reduced with the KLIP algorithm (see text). The contrast of the PSF in this image is $7.24\times10^{-7}$, which represents a 24 kW laser. \textit{Left}: a modeled GPI image with an injected, monochromatic laser pulse. \textit{Right}: the modeled spectrum of a 800 K planet (blue), a modeled 5000 K star (red), and a modeled 24 kW monochromatic laser (black).}
    \label{fig:monochromatic_laser}
\end{figure*}

Of course, such observations would have to detect a laser OIRSETI signal from an actual planet orbiting the target star. Potentially, this could be done in the time or spectral domains. A typical high-contrast imaging sequence includes multiple short exposures; if a laser source is variable on minute timescales, it would be clear through time-domain analysis. If not, any signal (planetary or otherwise) will likely be re-observed over months or years to determine its properties. If the source varies on timescales of hours or days (e.g. due to rotation of the source planet or orbital motion of a transmitter), it might be present in one epoch but not in another, though a true planet could also become unobservable if its orbit takes it inside the coronagraph’s inner working angle, requiring multiple observations. Spectroscopically, even the low-resolution spectrographs typically used for high-contrast imaging could distinguish between a narrowband laser source and a planet (see Figure~\ref{fig:monochromatic_laser}).

\section{Instrument Overviews} \label{subsec:introduction}
Here we present a brief overview of GPI and WFIRST and their data reduction processes. Table~\ref{tab:inst_summary} summarizes the relevant instrument details for GPI and WFIRST.

\begin{table}[h!]
\tablenum{2}
\centering
\caption{GPI \& WFIRST Instrument Details} \label{tab:inst_summary}
\begin{tabular}{ccc}
\hline
Property & GPI  & WFIRST \\
\hline
\hline
Wavelength              & 1-2.4 $\mu$m  & 546-604 nm  \\
Spectral Resolution (Band)          & 40 ($H$)  & 50 ($V$)  \\
Plate Scale ($^{\prime\prime}$/px)  & 0.014     & 0.0208  \\
Field of View ($^{\prime\prime}$)   & 5.6       & 10     \\
Number of Images                    & 40        & \nodata  \\
Exposure Time (s)                   & 60        & variable \\
\hline
\end{tabular}
\tablecomments{\citet{Macintosh+14, Kasdin+14, Kasdin+19, Larkin+14}}
\end{table}

\subsection{GPI Instrument Background}
GPI was designed to search for young, self-luminous giant planets that are heated by gravitational contraction. Typical target stars are less than 100 Myr old and located within \(\approx\) 100 pc of Earth; planets with masses above about 2 times that of Jupiter are detectable \citep{Macintosh+15}. 
	
GPI's adaptive optics system uses an I-band wavefront sensor to measure atmospheric turbulence and provide information for wavefront correction. The wavefront measurement accuracy, and hence overall GPI performance, is therefore determined by the I-band magnitude of the target star. The faint magnitude limit is \(\approx\) 10.0 mag due to detector readout noise \citep{Bailey_2016}. GPI has an apodized-pupil lyot coronagraph, with an inner working angle of 0.15$^{\prime\prime}$. Its primary science instrument is an an integral field spectrograph (IFS). A micro-lens array segments the field of view and a prism disperses light onto a HAWAII 2-RG detector at wavelengths in the Y, J, H, and K bands, where the K-band is split into two filters \citep[K1 and K2-bands,][]{wang_2017}. The spectral prism can be swapped for a Wollaston prism to observe in a polarimetry mode. GPI is usually employed in the H-band (1.5-1.8 \micron), so in this work we consider only that wavelength range.

\subsection{GPI Data Reduction}
GPI observations of a typical star consist of 40 individual 60-second exposures. The data are processed by the GPI Data Reduction Pipeline and the GPI Exoplanet Survey data analysis architecture \citep[DRP;][]{wang_2017}. The data reduction process can be broken up into three steps: data cube construction, stellar point spread-function (PSF) subtraction, and contrast curve generation. The raw IFS data consists of {$\sim$}35\,000 micro-spectra. The micro-spectra from each individual exposure are extracted from each image to construct a data cube with two spatial dimensions and one spectral dimension.  Even after GPI's coronagraph, the images of bright stars are dominated by residual diffracted starlight. The next step is obtaining a PSF subtracted data cube by angular differential imaging (ADI) and spectral differential imaging (SDI) in order to differentiate planet signal from the residual stellar PSF. Images at different paralactic angle rotation (ADI) or wavelengths (SDI) are used to generate a model of the stellar PSF, or speckle pattern, which is subsequently subtracted from the science images to recover the light of the planet.  In the simplest case, this model would just be an appropriately-scaled average of all the individual rotations or wavelengths. 

pyKLIP, a Python wrapper which implements the Karhunen-Loeve Eigenimage processing (KLIP) algorithm, allows for more sophisticated subtractions. KLIP creates an orthogonal basis of eigenimages and projects them to onto a science image \citep{Soummer_2012}. Typically for a given image or wavelength, a ``library" of other images is constructed to exclude images that overlap too much with the planet signal, i.e. in ADI images where the sky position of the planet has not rotated sufficiently or in SDI wavelength channels where the planet signal has not shifted or changed. A template of the expected planet spectrum is used, and defines which set of images can be used to construct the basis functions.  However, the more images that are excluded, the worse the fit to the PSF will be, as the PSF varies with wavelength. If the target planet spectrum is similar to that of the star and speckles, fewer images can be included in this library. When the target spectrum is a monochromatic laser, more images can be used to produce a better model of the PSF.

\begin{figure}[ht!]
\plotone{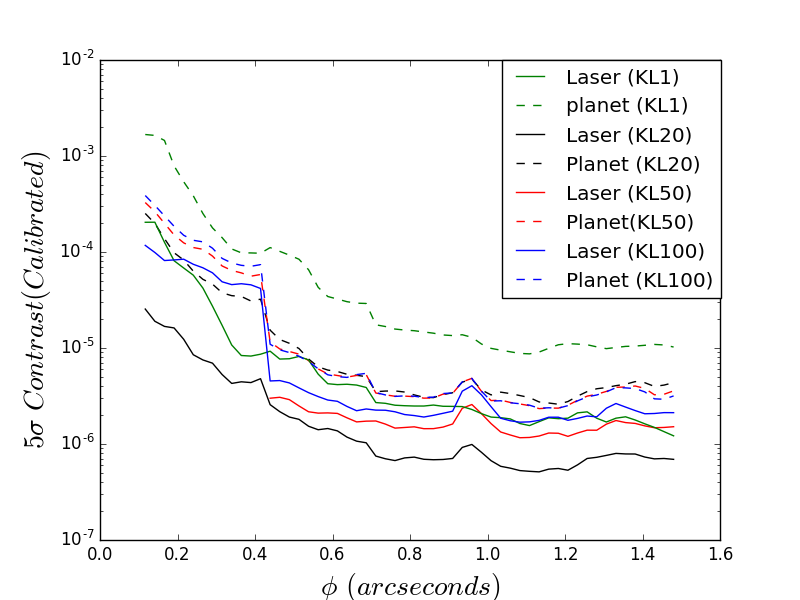}
\caption{Comparison of contrast for detection of monochromatic (e.g., laser) and broadband (e.g., planet) sources using different numbers of KL modes. }\label{fig:contrast_diff}
\end{figure}

To demonstrate this effect, a nearly monochromatic spectrum is injected into an existing data cube as shown in Figure~\ref{fig:monochromatic_laser}. The peak wavelength is 1.55 \(\micron\) and the full width at half maximum is 0.013 \(\micron\). The Gaussian profile of the spectra extracted from the data cube is also shown in Figure~\ref{fig:monochromatic_laser}. The monochromatic spectrum is plotted against a planet and stellar spectrum to demonstrate the difference between the two cases, and we evaluated the contrast for both classes of targets. The improvement in contrast between a continuous and monochromatic PSF is demonstrated in Figure~\ref{fig:contrast_diff}. In the typical case (using 20 KL modes), the signal to noise ratio of for detection of a monochromatic source is 30\% higher than the signal to noise ratio for detection of a broadband continuum source.

\subsection{GPI Contrast Curves}
The sensitivity of a high-contrast imaging instrument is usually expressed in terms of ``contrast," or the flux ratio of the faintest detectable planet to the residual scattered starlight. The final contrast is defined as five times the standard deviation of the noise (5$\sigma$) and expressed in terms of the planet to star flux ratio. The standard deviation $\sigma$ is calculated in concentric annuli as a function of separation. After KLIP processing, some of the signal of the planet has been subtracted out. In order to calibrate the true flux value of a planet in a reduced image, simulated planet spectra are injected into the original raw data cubes. These simulated images are then PSF subtracted in the same manner as the original observations, and the residual planet flux is evaluated \citep{Ruffio_2017}.

\subsection{WFIRST Overview}
 WFIRST is a space telescope planned for launch in 2025. Although its primary science capabilities are wide-field near-infrared imaging, it carries a prototype space coronagraph instrument, CGI \citep{Kasdin+14,Kasdin+19}. CGI uses coronagraph masks and two deformable mirrors to achieve contrast ratios of $\sim 10^{-9}$ and is intended to demonstrate the technologies for future exoplanet-imaging missions such as HABEX or LUVOIR \citep{Gaudi_2018, LUVOIRTeam_2018}, but will also be capable of imaging mature giant planets orbiting the nearest stars. Unlike GPI, CGI is most sensitive to wavelength ranges from 0.546-0.604 \micron. Operating at more sensitive contrast levels than GPI, it is potentially much more sensitive to faint laser transmitters. We evaluate the sensitivity of WFIRST in OIRSETI to show the potential of this technique. Future missions capable of detecting Earth-like planets will be even more sensitive, especially in the habitable zones of nearby stars. WFIRST carries both a direct imaging camera and (potentially) a restricted spectroscopic mode. We have evaluated CGI’s capabilities using only its direct imager, which would not produce any spectral information. However, in the event of a detected signal, the IFS could be used to distinguish a laser transmitter from a planet.

As with GPI, we evaluated the minimum detectable power limit of WFIRST by using the detectable flux ratios. To sample a range of stars, we simulated the detectable flux ratio of several targets with the EXOSIMS library V1.35 \citep{Savransky_2016} as function of angular separation. Table~\ref{tab:exosims} lists the arguments used in our simulations. To simulate a snapshot survey, we adjusted the integration time to approximately 60 minutes for each observation. We used a .JSON file (provided by D. Savransky, private communication) of the updated instrument specifications in our simulations. This simulation was executed by a single function which calculated each planet's $\Delta$magnitude limit given an integration time. The magnitude difference limit was evaluated at a range of angular separations. The $\Delta$magnitudes were converted into a flux ratio proportional to the contrast defined in previous paragraphs.

\begin{deluxetable}{cc}
\tablecaption{Arguments Used in WFIRST Simulations}
\tablecolumns{2}
\tablenum{1}\label{tab:exosims}
\tablehead{
\colhead{Argument} & \colhead{Value} }
\startdata
Integration Time (days)     & 0.06 \\
Target Lists                & EXOCAT1 \\
Star Index                  & HIP name of target star\\
Zodiacal light              & stark \\
Working Angle ($^{\prime\prime}$)   & 0.15-0.3 \\
Mode                        & Imager \\
\enddata
\end{deluxetable}

\section{Method} 
\subsection{Target Stars \label{subsec:gpicurves}}
\begin{figure*}[ht!]
\plotone{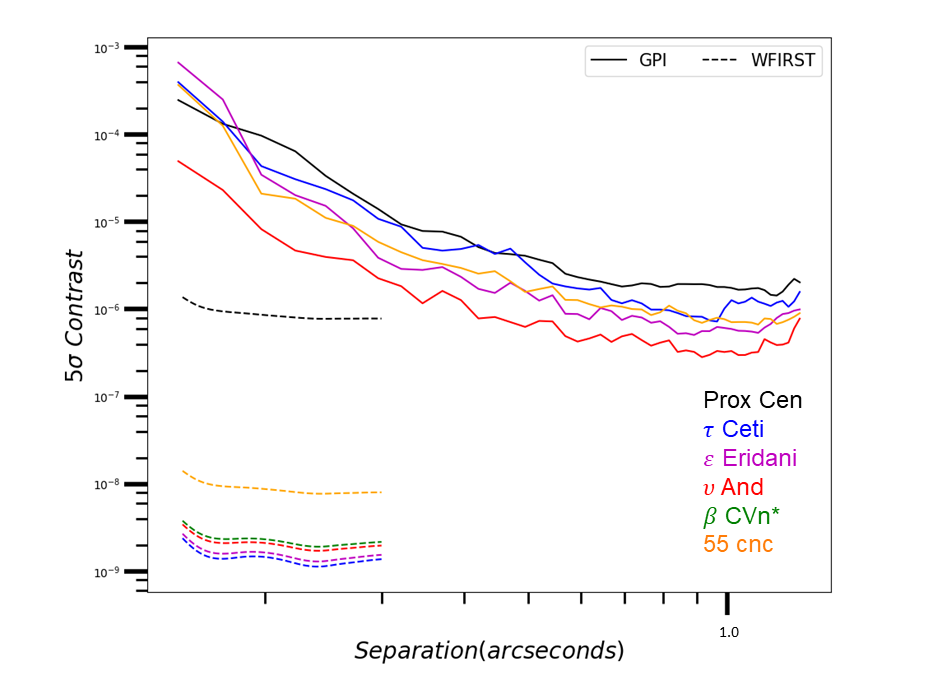}
\caption{The contrast curves for all analog and simulated stars. Note that for the GPI case $\beta$ CVn and $\upsilon$ And share the same analog star as they have similar I-band magnitudes. }\label{fig:contrast_curves}
\end{figure*}

We use 6 target stars for both GPI and WFIRST. The target stars are a sample of each F or later spectral type (with an emphasis on G stars). The target stars are on the main sequence, host one or more known or candidate  exoplanets (with the exception of $\beta$ CVn, discussed further below), are within 20 parsecs, and  have an  $M_{\rm I}\lesssim$ 9.5. The G stars were chosen based off their potential for habitability and to demonstrate variability in distances. The target stars and their properties are summarized in Table~\ref{tab:target_stars}; the 5$\sigma$ measured or estimated contrast curves for each instrument on these stars is shown in Figure~\ref{fig:contrast_curves}. For GPI, these contrast levels typically correspond to the near-infrared emission from a planet with a mass of several Jupiter masses; for WFIRST, to a planet with a radius similar to Jupiter's at a physical separation of 3-5 AU.

\begin{table*}
\tablenum{3}
\centering
\caption{Summary of Target Stars} \label{tab:target_stars}
\begin{tabular}{cccccc}
\hline
Star & Spectral Type$^{a}$  & Distance (pc)   & I magnitude & V magnitude &\# Exoplanets$^b$ \\
\hline
\hline
$\nu$ And   & F8V       & 13.49 & 3.35 &  4.10  & 3 \\
$\beta$ CVn   & G0V    & 8.43&  3.42 &  4.25  & 0 \\
$\tau$ Ceti & G8V       & 3.65 &  2.41 & 3.50 & 4 \\
55 Cnc      & G8V       & 12.59& 5.0& 5.95 & 5 \\
$\epsilon$ Eri  & K2V   & 3.22 & 2.54 &  3.73 & 1 \\
Proxima Cen & M5.5Ve    & 1.30&  7.41  &  11.13 & 1 \\
\hline
\end{tabular}
\tablecomments{$^a$Taken from the SIMBAD database. $^b$Confirmed or candidate exoplanets from The Exoplanet Orbits Database.}
\end{table*}

\subsection{The Laser Model}\label{sec:laser_model}
Our results assume a case where ETI would transmit a laser of wavelength \(\lambda\) centered in GPI's H-band (1.55 \(\micron\)) or in WFIRST's V-band (575 nm). These lasers would be transmitted from a diffraction limited telescope with a diameter $D$ \citep[assumed to be 10 m, as is standard in the OIRSETI calculations; e.g.,][]{Tellis, wright2014}.

The detectable power limit, $P$, can be scaled for any laser and transmitter combination. The detection limit can be scaled for a transmitter of any size as $P\propto D^{-2}$. As the diameter of the transmitter increases the divergence angle $\theta$ of the beam decreases, thus yielding a lower detectable power limit. This relationship also allows for the evaluation of a variety of integration times, for example, as integration time increases (in the case of WFIRST, which could observe a target for multiple days) the sensitivity would increase by the square root of the exposure time. This relationship is valid under the assumption that the laser remains on its target for the entire integration period. Additionally,  $P\propto \lambda^{2}$, where \(\lambda\) must be within an instrument's filter bandwidth. This allows for scaling the results to GPI's J-band, where many commercial lasers are manufactured.

The flux density of the laser is modeled as a Gaussian profile. The Gaussian intensity at a single wavelength is calculated from the wavelength of the laser, \(\lambda\), and the diameter of the diffraction limited telescope. The divergence angle, \(\theta\), is therefore defined as

\begin{equation} \label{eq:4}
\theta= \frac{\lambda} {D} \approx  \frac {\lambda } {\pi w_0}.
\end{equation}

\noindent The beam waist radius, \( w_0\), is defined as the radius of the \(1/e^2\) contour when the wavefront is flat\footnote{See \url{http://www.cvimellesgriot.com/Products/Documents/TechnicalGuide/Laser-Guide.pdf}}. We use the divergence approximation to infer that \(  w_0 \approx D/\pi\). From this approximation of \( w_0\), the radius $r$ of the \(1/e^2\) contour when the wavefront has traveled some distance $z$ can be calculated as:

\begin{equation}
r(z)= \frac{z\lambda }{\pi w_0} \approx \frac{z\lambda}{D}.
\end{equation}

\begin{figure*}[ht!]
\plotone{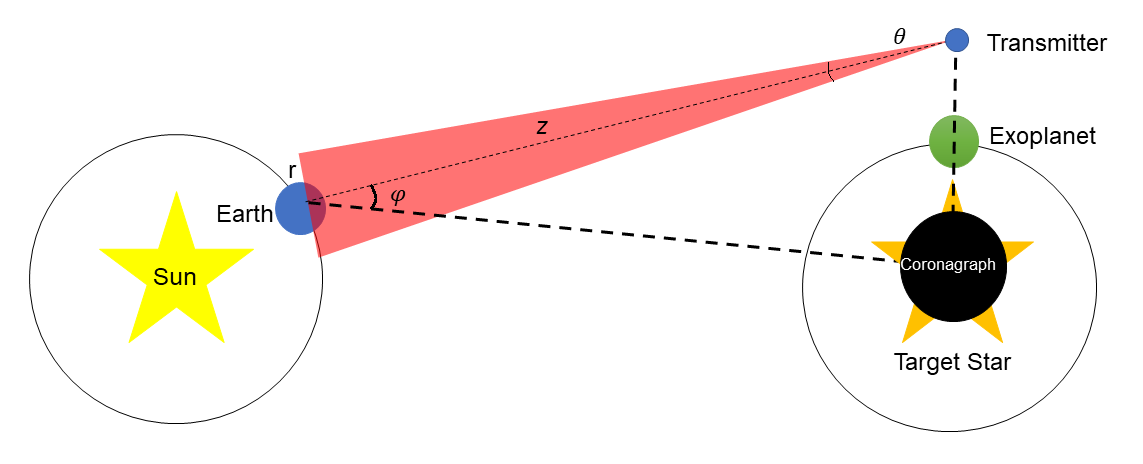}
\caption{A cartoon diagram of the OIRSETI geometry considered in this work}. A laser is broadcast from (or in orbit around) an exoplanet and aimed at Earth. The distance from the transmitter to Earth is denoted as $z$. The coronagraph blocks the starlight, enabling the detection of the planet (and laser signal). The parameters $r$, $\theta$, and $\phi$ are defined in the text (see Sections~\ref{sec:laser_model} and \ref{sec:results}).   \label{fig:laser_schematic}
\end{figure*}

\noindent The geometry of an incoming OIRSETI laser signal, as observed by an Earth-observer, is shown in Figure~\ref{fig:laser_schematic}.

With a 10-m transmitter telescope, this angle \(\theta\) is quite small -- 0.01$^{\prime\prime}$ for visible wavelengths, corresponding to 0.1 astronomical units at a distance of 10 pc. For a laser beam to be detectable on Earth it must therefore be carefully aimed. For a transmitter 10 pc away this requires knowledge of the proper motion of our sun at the level of $\sim30$ microarcseconds per year (easily obtained with a Gaia-like mission) and knowledge of the orbital elements of the Earth to within a few percent (requiring an ET space coronagraph.) Alternatively, if 1-m transmitter is used with a correspondingly broader beam, less precision will be required.  

\section{Results}\label{sec:results}
Currently, GPI has not observed any of the nearby target stars except for $\varepsilon$ Eridani. Before the detection limit for each target star can be calculated, an ``analog'' star must be found. An analog star is any star imaged with the IFS by the GPI Exoplanet Survey \citep[GPIES;][]{Macintosh_2019} with an I-band magnitude matching that of the target star. The final contrast curve for the analog star then determines the detectable flux density of the laser around the target star (this does not apply to the WFIRST calculations, where we instead use simulated contrast curve).

The final contrast curve of the analog star is converted from contrast as a function of angular separation to a magnitude difference as a function of angular separation. The flux density of the target star, \(F_T\), is then converted from magnitudes to units of W m$^{-2}$ \micron$^{-1}$ (using the flux density of an A0 star in the H-band as a reference value). The flux density of the laser, \(F_L\), is then calculated as the minimum flux density that would be detectable for a given contrast curve. The Gaussian optics approximations (discussed in the previous section) are applied to obtain the value of the width of the beam when it reaches the detector, $w(z)$, which is dependent on the distance $z$ to the target star. Assuming the laser intensity follows a Gaussian distribution in angular separation $\phi$, we can calculate the detectable power limit $P(\phi)$ over the bandwidth of the filter, \(\Delta\lambda\), by:

\begin{equation}\label{eq:5}
P(\phi) = F_{l}(\phi)\pi r(z)^2\Delta\lambda,
\end{equation}

\noindent where we have assumed the a rectangular filter bandpass.

\begin{figure*}[ht!]
\plotone{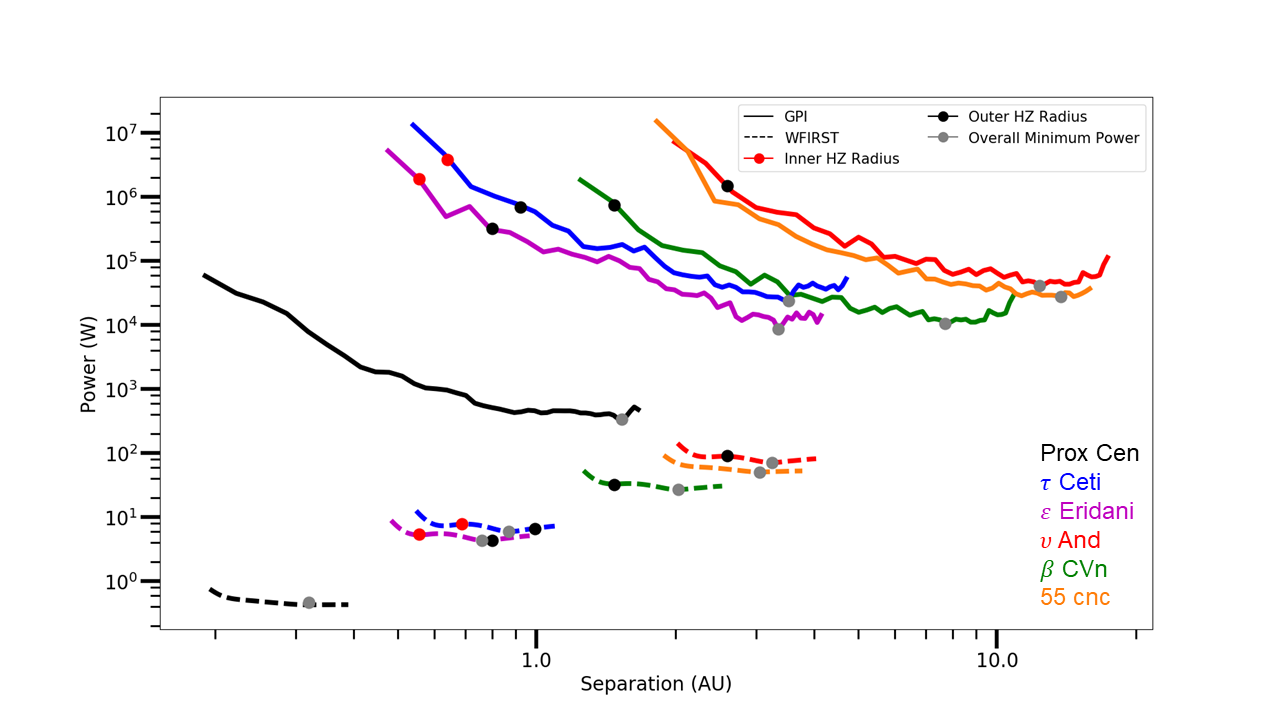}
\caption{Laser detection thresholds for GPI (solid lines) and WFIRST (dashed lines) for eight analog stars. If applicable, the HZ region is indicated (red dots denote $r_{\rm in}$ and black dots denote $r_{\rm out}$). The solid gray dots indicate the overall minimum power for each curve. The minimum power within the HZ lies at the outer edge for all GPI cases and two WFIRST cases (\(\beta\) CVn and \(v\) And), or shares a point with the overall power minimum. Power would scale as 100 m$^2$/$D^2$ for smaller telescopes.}\label{fig:results}
\end{figure*}

Figure~\ref{fig:results} shows a ``cartoon'' diagram illustrating the geometry of the laser signal aimed at Earth from a nearby exoplanet, with the parameters $r$, $z$, $\phi$, and $\theta$ indicated on the diagram.

Figure~\ref{fig:results} shows the minimum power of an extraterrestrial laser (as a function of distance from the parent star) that would be detectable by GPI and WFIRST for each of the target stars in our study. For each target star, we calculate the radius of the laser beam when it reaches the detector ($r_{\rm beam}$), the overall minimum detectable laser power $P_{\rm min}$, and the separation $r_{\rm min}$ from the parent star at which $P_{\rm min}$ would be detected (we assume a launch telescope with $D=10$ m). We additionally estimate the inner and outer radii (in AU) of each star's HZ: the inner radius $r_{\rm in}$ is estimated as $(L_{\odot}/1.1)^{1/2}$, and the outer radius $r_{\rm out}$ is given by $(L_{\odot}/0.53)^{1/2}$ \citep{Kasting1993}. For stars where the HZ is resolved by either GPI or WFIRST, we additionally measure the minimum laser power ($P_{\rm HZ,min}$) detectable at a distance within the HZ ($r_{\rm HZ,min}$). Our results are summarized in Table~\ref{table:results}. We note that one target star, $\tau$ Ceti, hosts four exoplanet candidates, two of which fall within the star's HZ \citep{Feng+17}. Our analysis shows that both GPI and WFIRST are able to spatially resolve the entirety of the $\tau$ Ceti HZ.

$\beta$ CVn is another particularly interesting target star since it is the closest solar analog and is part of the Catalog of Habitable Stellar Systems (HabCat). Despite the lack of known exoplanets, this star is still of considerable interest due to its age, size, metallicity, variability, kinematics, and spectral type \citep{Turnbull_2003}. The outer radius of $\beta$ CVn's HZ is resolvable by both GPI and WFIRST. 

\begin{table*}
\tablenum{4}
\scriptsize
\centering
\caption{Minimum Laser Power Thresholds in GPI and WFIRST} \label{table:results}
\begin{tabular}{cccccccccccccc}
\hline
        & \multicolumn{6}{c}{GPI}   && \multicolumn{6}{c}{WFIRST}   \\ \cline{2-7} \cline{9-14}
Star    & $r_{\rm beam}$   & $r_{\rm min}$ & $P_{\rm min}$ & $H_{\rm min}$ & $r_{\rm HZ,min}$   & $P_{\rm HZ,min}$  &&   $r_{\rm beam}$   & $r_{\rm min}$ & $P_{\rm min}$& $V_{\rm min}$ & $r_{\rm HZ,min}$   & $P_{\rm HZ,min}$  \\
        & (AU) & (AU)  & (kW) & (mag) & (AU)  & (kW)  && (AU) & (AU)  & (W)  & (mag) & (AU)  & (W)   \\
\hline
\hline

$\nu$ And & 0.22 &  12.36 & 40.70 & 16.37 & 2.58& 1519.48 &&0.08&3.24&70.67& 21.91&2.31& 86.55\\ 

$\beta$ CVn & 0.06 & 7.72  & 10.40  & 16.37& 1.46 & 778.77&& 0.05 & 2.03 &  26.76& 21.80  & 1.46 & 32.81   \\

$\tau$ Ceti & 0.06  & 3.53  & 23.90  & 15.35 & 0.92 & 741.15 &&  0.02   & 0.87 & 5.90& 22.37   & 0.87 & 5.90  \\

55 Cnc  & 0.20 &    13.77&     27.24&15.44&\nodata&\nodata&&0.07&3.05&50.39&20.28&\nodata&\nodata\\

$\epsilon$ Eri & 0.05 &3.35 &8.60&15.74&0.80& 344.05 && 0.02 & 0.76&4.27& 22.23 &0.76& 4.27\\

Proxima Cen  & 0.02&1.53 &0.336&14.61 &\nodata&\nodata & & 0.01 & 0.32 &0.46&15.35&\nodata&\nodata\\
\hline
\end{tabular}
\end{table*}

The power outputs demonstrated here are well within our own current technological abilities. Since the 1980's, MW class lasers have been in use for military purposes. One example is the Mid-Infrared Advanced Chemical Laser (MIRCAL), a CW laser capable of output of over a MW for up to 70 seconds. This remains the world's highest power CW laser\footnote{\url{http://helstf-www.wsmr.army.mil/miracl.htm}}, and would be well within the detection threshold of GPI if coupled to a 1-m telescope. The Deep Space Optical Communication experiment\footnote{\url{https://en.wikipedia.org/wiki/Deep_Space_Optical_Communications}} uses a 5 kW laser on a 1-m telescope and would be detectable by WFIRST for most of our target stars. 

A comparison to other detection methods of optical and infrared laser signals shows GPI and WFIRST have unique capabilities. Many studies \citep[e.g.,][]{Howard+04} have been limited to pulsed lasers. Coronagraphic imaging instruments are sensitive to both CW and pulsed lasers. \citet{Tellis+15} show that spatially resolved Keck HIRES spectroscopy searching for narrowband sources along the HIRES slit would achieve detection limits of 90 W from \(\approx\) 30 pc and 1kW at \(\approx\) 300 pc, assuming narrow-band signals from a 10 m diffraction limited transmitter. While this is a deeper detection limit than GPI, their search is limited to lasers centered at 364-789 nm at angular separations of 2-7$^{\prime\prime}$.  This smaller wavelength range condenses the power considerably, allowing for a lower detection limit at larger interstellar distances. However, their inner working angle (2000 AU) is far from the habitable zone of a Sun-like star. \citet{Tellis} additionally established detection limits using unresolved Keck HIRES spectroscopy within the same wavelength channel (i.e. mixed with the stellar spectrum) ranging from 3kW to 13 MW. This was roughly independent of distance, since they varied the exposure times for stellar spectra in order to yield a nearly consistent number of photons per pixel, and therefore they achieved similar detection limits relative to the stellar continuum for any laser emission line. Our detection limit results for GPI and WFIRST are below the \citet{Tellis+15,Tellis} limits, and comparable to some commercially available and higher powered lasers. The geometry in Figure~\ref{fig:laser_schematic} can also be reversed to establish the minimum power required for an Earth-based laser to outshine the Sun. This has been done for several test cases recently \citep[see, e.g.][]{Clark+18} for lasers with wavelengths of 1.315 \micron\ (J-band), 1.064 \micron, and 785 nm. Their transmitters considered diameters of future generation class observatories. 

As discussed above, this analysis assumes a large space-based transmitter. The main limitation of a ground-based transmitter is atmospheric turbulence; if not equipped with sophisticated adaptive optics, the outgoing beam would be spread over a much larger angle. The typical FWHM of such a beam is \(\lambda/r_{0}\), where \(r_{0}\) is the Fried seeing parameter. The detectable power could therefore be calculated by scaling the powers calculated here for a 10-m telescope by the ratio of $(10 \text{\rm m}/r_{0})^2$. For a typical astronomical site, \(r_{0}\) at 500 nm wavelength is about 10 cm, and at 1.6 $\mu$m is about 40 cm. The detectable power for a star such as $\tau$ Ceti would be 15,000 kW  for GPI and 60,000 W for WFIRST. 

The near-IR has long been expected to be the best wavelength in which to search for ET \citep{Townes+61,Townes83}, however searches of this type are not often done. This work presents an example of near-IR SETI, and establishes the minimum power of laser we would need if ET had GPI or WFIRST-class detection instruments and could accommodate currently existing transmitters.

\section{Conclusion}
This work evaluates the detection limits of coronagraphic imaging instruments to optical and infrared lasers to determine the effectiveness of GPI and WFIRST at detecting such ETI technosignatures. We examined the GPI data reduction pipeline and PSF algorithms, and showed that it yields a  30\% higher S/N when detecting a monochromatic signals compared to a broadband signal. We then modeled a laser with a Gaussian intensity distribution and used archival data to simulate the detection limits of GPI. We also modeled a sample of contrast curves using the EXOSIMS library to demonstrate the increase in performance that the next generation of space telescopes (e.g., WFIRST) will provide. Our target stars were based on Sun-like, planet-hosting stars within 20 pc of Earth.  The most notable result was the detection limit for \(\tau\) Ceti, a G5V star with four known exoplanets, two of those within the habitable zone (HZ). The result shows that a 24 kW laser is detectable from
\(\tau\) Ceti from outside of the HZ with GPI and a 7.3 W laser is detectable from within \(\tau\) Ceti’s HZ by WFIRST. We discussed how these results compare to past studies and how the power of the lasers fit into current technology. 

This is a novel method of modeling the detection limits of coronagraphic exoplanet imaging instruments with results that can be scaled for any transmitter/laser configuration, since power is proportional to the squared inverse of the transmitter size and the wavelength of the laser squared. Our method successfully shows the application of CORSETI.

\acknowledgements
The authors would like to thank M. S. Povich and S. Wright for useful discussions.

This work uses observations obtained at the Gemini Observatory, which is operated by the Association of Universities for Research in Astronomy, Inc., under a cooperative agreement with the NSF on behalf of the Gemini partnership: the National Science Foundation (United States), the National Research Council (Canada), CONICYT (Chile), Ministerio de Ciencia, Tecnolog\'{i}a e Innovaci\'{o}n Productiva (Argentina), and Minist\'{e}rio da Ci\^{e}ncia, Tecnologia e Inova\c{c}\~{a}o (Brazil). This research has made use of the SIMBAD and VizieR databases, operated at CDS, Strasbourg, France. The Gemini Planet Imager Exoplanet Survey was supported by Supported by NSF grants AST-1411868 and AST-1518332 and NASA grants NNX14AJ80G, NNX15AC89G and NNX15AD95G and NN15AB52l. This work benefited from NASA's Nexus for Exoplanet System Science (NExSS) research coordination network sponsored by NASA's Science Mission Directorate.


\end{document}